\documentclass[usenatbib]{mnras}

\pdfoutput=1
\usepackage{amssymb,amsmath}
\usepackage{graphicx}
\usepackage{multirow}
\usepackage{natbib}
\usepackage{tabularx}
\usepackage{footmisc}
\usepackage{wasysym}
\usepackage{upgreek}

\usepackage{color}
\title{Fast Radio Transient searches with UTMOST at 843 MHz}

\author[M.~Caleb et al.]
{M.~Caleb$^{1,2,3}$\thanks{Email: manisha.caleb@anu.edu.au},
C.~Flynn$^{2,3}$,
M.~Bailes$^{2,3}$,
E.\,D.~Barr$^{2,3}$,
T.~Bateman$^{5}$,
S.~Bhandari$^{2,3}$,
\newauthor D.~Campbell-Wilson$^{6,3}$,
A.\,J.~Green$^{6,3}$,
R.\,W.~Hunstead$^{6}$,
A.~Jameson$^{2,3}$,
F.~Jankowski$^{2,3}$,
\newauthor E.\,F.~Keane$^{4,2,3}$, 
V.~Ravi$^{2}$,
W.~van Straten$^{2}$,
V.~Venkataraman Krishnan$^{2,3}$
\\ \\
$^{1}$ Research School of Astronomy and Astrophysics, Australian National University, ACT, 2611, Australia\\
$^{2}$ Centre for Astrophysics and Supercomputing, Swinburne University of Technology, P.O. Box 218, Hawthorn, VIC 3122, Australia \\
$^{3}$ ARC Centre of Excellence for All-sky Astrophysics (CAASTRO)\\
$^{4}$ SKA Organisation, Jodrell Bank Observatory, Cheshire, SK11 9DL, UK\\
$^{5}$ CSIRO Astronomy \& Space Science, Australia Telescope National Facility, P.O. Box 76, Epping, NSW 1710, Australia\\
$^{6}$ Sydney Institute for Astronomy (SIfA), School of Physics, The University of Sydney, NSW 2006, Australia}

\begin{document}


\maketitle


\begin{abstract}
\noindent We report the first radio interferometric search at 843 MHz for fast transients, particularly Fast Radio Bursts (FRBs). The recently recommissioned Swinburne University of Technology's digital backend for
the Molonglo Observatory Synthesis Telescope array (the UTMOST) with its large collecting area (18,000
$\mathrm{m^2}$) and wide instantaneous field of view (7.80 $\mathrm{deg^2}$) is expected to be an efficient tool
to detect FRBs. As an interferometer it will be capable of discerning whether the FRBs are
truly a celestial population. We show that UTMOST at full design sensitivity can 
detect an event approximately every few days. We report on 2 preliminary FRB surveys at about 7\% and 14\% respectively of the array's final sensitivity. Several pulsars have been detected via single pulses and no FRBs were discovered with pulse widths ($W$), in the range 655.36 $\upmu$s $< W < 41.9$ ms and dispersion measures (DMs) in the range $100 < $DM$< 2000$ $\mathrm{pc\,cm^{-3}}$. This non-detection sets a 2$\sigma$ upper limit of the sky rate of not more than 1000 events $\mathrm{sky^{-1}}$ $\mathrm{day^{-1}}$ at 843 MHz down to a flux 
limit of 11 Jy for 1 ms FRBs. We show that this limit is consistent with previous survey limits at 1.4 GHz and 145 MHz and set a lower limit on the mean spectral index of FRBs of $\alpha > -3.2$.
\end{abstract}

\begin{keywords}
instrumentation: interferometers -- pulsars: general -- intergalactic medium -- surveys -- methods: data analysis   
\end{keywords}

\section{Introduction}

High time resolution astronomy over the last decade has led to the discovery of new classes of radio sources such as the Rotating Radio Transients (RRATs) \citep{Maura} and Fast Radio Bursts (FRBs) (\citeauthor{Lorimer} \citeyear{Lorimer}; \citeauthor{Thornton} \citeyear{Thornton}; \citeauthor{Spitler} \citeyear{Spitler}; \citeauthor{Spolaor-Bannister} \citeyear{Spolaor-Bannister}; \citeauthor{Petroff_FRB} \citeyear{Petroff_FRB}; \citeauthor{Ravi} \citeyear{Ravi}; \citeauthor{Champion} \citeyear{Champion}; \citeauthor{Masui} \citeyear{Masui} ; Keane et al. in prep). The majority of the RRATs and all the FRBs are characterized by millisecond duration pulses implying coherent physical processes in their origin if they are compact sources. RRATs have been found to repeat on timescales of a few pulses an hour to a few pulses a day while FRBs have not yet been seen to repeat.
These elusive bursts have $\sim$ Jy peak flux densities and dispersion measures that well exceed the contribution from the Milky Way along the line of sight indicative of a possible a cosmological origin. 
Only a handful of FRBs are known, and to date no transient event or afterglow has been seen at any other wavelength despite major efforts to do so \citep{Petroff_FRB}. Several cosmological and non-cosmological models for the origin of FRBs have been suggested, including radio emission from pulsars \citep{Wasserman, Connor}, collapsing gravitationally unstable black holes \citep{Falcke}, hyper flares from magnetars \citep{Lyubarsky} and dark matter induced collapse of neutron stars \citep{Fuller}.

Caleb et al. 2015 (MNRAS submitted) have performed Monte Carlo simulations of a cosmological population of FRBs to study the distributions of their observed and inferred properties and their log$N$-log$\mathcal{F}$ curve. 
From comparison of the slope of the log$N$-log$\mathcal{F}$ curves of their simulations with the slope of the log$N$-log$\mathcal{F}$ curve of the observations they conclude that FRBs are consistent with being of cosmological origin.
If FRBs are indeed cosmological in origin, they could be potentially used to probe the `missing baryon problem' \citep{McQuinn}, obtain rotation measure of the intergalactic medium (IGM) along the line of sight \citep{Zheng} and also as an independent measure of the dark energy equation of state \citep{Zhou}. 
Most of the FRBs that have been discovered to date i.e. between 2007 and 2015, have all been seen at 1.4 GHz using single dish antennas with relatively poor angular resolution. This means that their spatial localisation is poor and lacks the precision required to unequivocally associate them with a possible host galaxies. Until 2013, FRBs had only been discovered in archival surveys, but since 2014 we have entered the era of real time detections with rapid multi-wavelength follow-up with three already having been done at Parkes (\citeauthor{Petroff_FRB} \citeyear{Petroff_FRB}; Keane et al. in prep).
\cite{Thornton} estimate the FRB event rate as $1.0^{+0.6}_{-0.5}\times10^4$ $\mathrm{sky^{-1}}$ $\mathrm{day^{-1}}$. \cite{KeanePetroff} have reanalyzed the \cite{Thornton} results and derive a fluence complete event rate of 2500 events $\mathrm{sky^{-1}}$ $\mathrm{day^{-1}}$ above a fluence of 2 Jy ms. 

There is clearly a need to discover FRBs more efficiently as the present discovery rate is only of order 1 per $\sim$ 12 days on sky at Parkes.
The 50 year old Molonglo Observatory Synthesis Telescope in Australia is currently being refurbished with a new digital backend system and increased bandwidth as part of an upgrade to transform it into a burst finding machine. This instrument being an interferometer will help discern if FRBs are truly a celestial population by measuring a parallax to the sources. In this paper we introduce the Molonglo observatory synthesis telescope and discuss its single pulse sensitivity in Section \ref{sec:MOL}. The first FRB survey at 843 MHz using the UTMOST instrument and limits on the detectability of FRBs is discussed in Section \ref{sec:survey}.
We make estimates of the FRB rates we can expect with the UTMOST instrument, showing that at full sensitivity it is considerably more
effective than Parkes for doing FRB surveys due to its large field-of-view and high observing duty cycle (Section \ref{sec:MR}) under conservative assumptions for the FRB spectral index. At full sensitivity we expect to detect an event every few days. We constrain the FRB event rate and mean spectral index based on the non-detection of FRBs in these pilot surveys and draw our conclusions in Section \ref{sec:conc}.

\section{The Molonglo Observatory Synthesis Telescope (MOST)}
\label{sec:MOL}

The Molonglo telescope was originally a ``Mills Cross" design, completed 
in 1967 \citep{Mills} and operating as a transit instrument at 408 MHz. It is located about 300 km 
south-west of Sydney, near Canberra, and is a field station of the University of Sydney. It played a crucial role in radio astronomy with the discovery of the Vela pulsar \citep{Large} and 155 new pulsars in the second Molonglo pulsar survey \citep{Manchester}. It was substantially modified in the early 1980s to make the East-West (E-W) arms fully steerable and increase the operating frequency to 843 MHz and a 3 MHz bandwidth \citep{Robertson}. The telescope was renamed the Molonglo Observatory Synthesis Telescope (MOST); it has a collecting area of 18000 m$^2$, the largest in the Southern hemisphere.

The E-W arm consists of two collinear cylindrical paraboloids, each 11.6 m wide and 778 m long, separated by a 15 m gap \citep{Bock}. Each paraboloid is divided into smaller sections called ``modules'', each with a beam of order $4.64^\circ \times 2.14^\circ$
(EW-NS). Four such modules were linked together digitally to form a ``bay'' and 44 such bays constitute one arm. A
line feed system of 7744 right circularly polarised dipoles (22 per module), in 352 resonant chambers each feeding a Low Noise Amplifier
(LNA) means that the telescope is effectively an array of 352 receivers operating at a system temperature of $\sim$ 70 K \citep{Duncan}. The E-W arms can be tilted North-South (N-S), while E-W pointing is attained by differential rotation of the ring antennae (spaced at 0.54
$\lambda$) on the line feed. The telescope can access the whole sky south of $\delta = +18^{\circ}$, although hour angle coverage is
limited to an E-W tilt of $\pm 60^{\circ}$. 

The telescope is currently being upgraded both in the backend receivers and with the installation of a
new graphics processing unit (GPU) based correlator, in a collaboration between Sydney and
Swinburne Universities. The installation of high-performance  GPUs at MOST has transformed it into a powerful instrument, the Swinburne University of Technology upgrade for the MOST (UTMOST ; Bailes et al. in prep) and enlarged the field of view to twice that of the Sydney University Molonglo Sky Survey \cite[SUMSS ;][]{Bock} due to processing data from each `module' rather than each `bay'. 




\begin{table} 
\caption{Comparison of Parkes multibeam \citep{Manchester} and UTMOST (Bailes et al. in prep)} 
\centering
\label{tab:specs}
\begin{tabular}{c c c}
\hline\hline
Parameter                   & Parkes & UTMOST \\ [0.5ex] 
\hline 
Field of View ($\mathrm{deg^2}$) & 0.55  & $4.64\times2.14$   \\
Central beam Gain (K $\mathrm{Jy^{-1}}$) & 0.7   & 3.6 \\
Central beam $T_\mathrm{sys}$ (K)       & 21    & 70  \\ 
Frequency (MHz)             & 1352  & 843 \\
Bandwidth (MHz)             & 340   & 31.25  \\
Channel width (kHz)         & 390.625 & 781.25 \\
No. of polarisations        & 2     & 1   \\
Polarisation feeds        & Dual linear   & Right circular  \\ 
Fresnel limit (km) & $\sim$ 40 & $\sim$ 14,000 \\[1ex] 
\hline  
\end{tabular} 
\end{table}

The sensitivity of UTMOST to FRB events (i.e. single pulse events) can be
calculated using the radiometer equation,
\begin{equation}
S_\mathrm{min} = \beta \, \frac {\mathrm{S/N} \, (T_\mathrm{rec} + T_\mathrm{sky})} {G \, {\sqrt{B \, t \, N_\mathrm{p}}}}
\end{equation}
\noindent where ${S_\mathrm{min}}$ is the minimum detectable flux for a given signal to
noise (S/N), $\beta$ is the digitisation factor, 
$B$ is the bandwidth in Hz, ${N_\mathrm{p}}$ is the number of
polarisations, $t$ is the width in seconds, ${T_\mathrm{rec}}$ and ${T_\mathrm{sky}}$ are the receiver and sky
temperatures in K respectively, and $G$ is the system gain in
K Jy$^{-1}$. 
For a pulse with S/N of 10 and width of 1 ms at Parkes, the sensitivity is ${S_\mathrm{min}} =$ 0.4 Jy. At UTMOST, ${S_\mathrm{min}} =$ 1.6 Jy. Thus UTMOST is about four times less sensitive to individual FRB events than Parkes. This is more than compensated for with its 14 times larger field-of-view indicating that UTMOST can be a very effective FRB discovery machine. In practice the sensitivity at UTMOST degrades depending on the scattering from the ISM and possibly the IGM at the lower UTMOST operating frequency. 
Detailed calculations of the event rate at UTMOST, taking into account the system sensitivity, sky temperature at our operating frequency, scattering effects due to the ISM and IGM, DM smearing due to channel bandwidth, beam pattern of the telescope and adopted FRB co-moving space density have been performed in a companion paper, Caleb at al. 2015 (MNRAS submitted). 

The main properties of UTMOST and Parkes from the point of view of
discovering FRBs are shown in Table \ref{tab:specs}.

\section{FRB surveys at UTMOST}
\label{sec:survey}

Two FRB searches have been performed at UTMOST at different fractional sensitivities during the ongoing upgrade. These two surveys are called V1.0 and V2.0. The antennae are aligned and fringe stopped to maintain stable and flat phases and then combined
into a tied-array beam, centered on the primary beam boresight. This
beam is then re-steered into 352 tied array beams called ``fan beams'' that are ``tiled''
across the 4 degree East-West axis of the primary beam. Time series
from each fan beam are detected and integrated from 1.28 to 655.36 $\upmu$s
sampling and also requantised to 8-bits/sample. 

The total data rate to the backend is 11 GBps, and the resulting output data
rate from the 352 beams is approximately 10 MBps for both surveys. The input stream at UTMOST in FRB search mode is 16 MHz for survey V1.0 and 31.25 MHz for V2.0, of single polarisation baseband data from 352
antennae, in 20 frequency channels produced in a polyphase filterbank
(PFB). We upgraded to 40 coarse channels in FRB survey V2.0. 
After completion of V1.0, the rest of the GPUs were installed onsite (May 2015) so that the full 31.25 MHz could be processed for V2.0. As a consequence it is clear that roll-off at the edge of the bandpass is quite pronounced so that the extra bandwidth is not usable. We conservatively assume 16 MHz of final effective bandpass for all the results in this paper.

\begin{figure*}
\centering
\includegraphics[width=5.7 in]{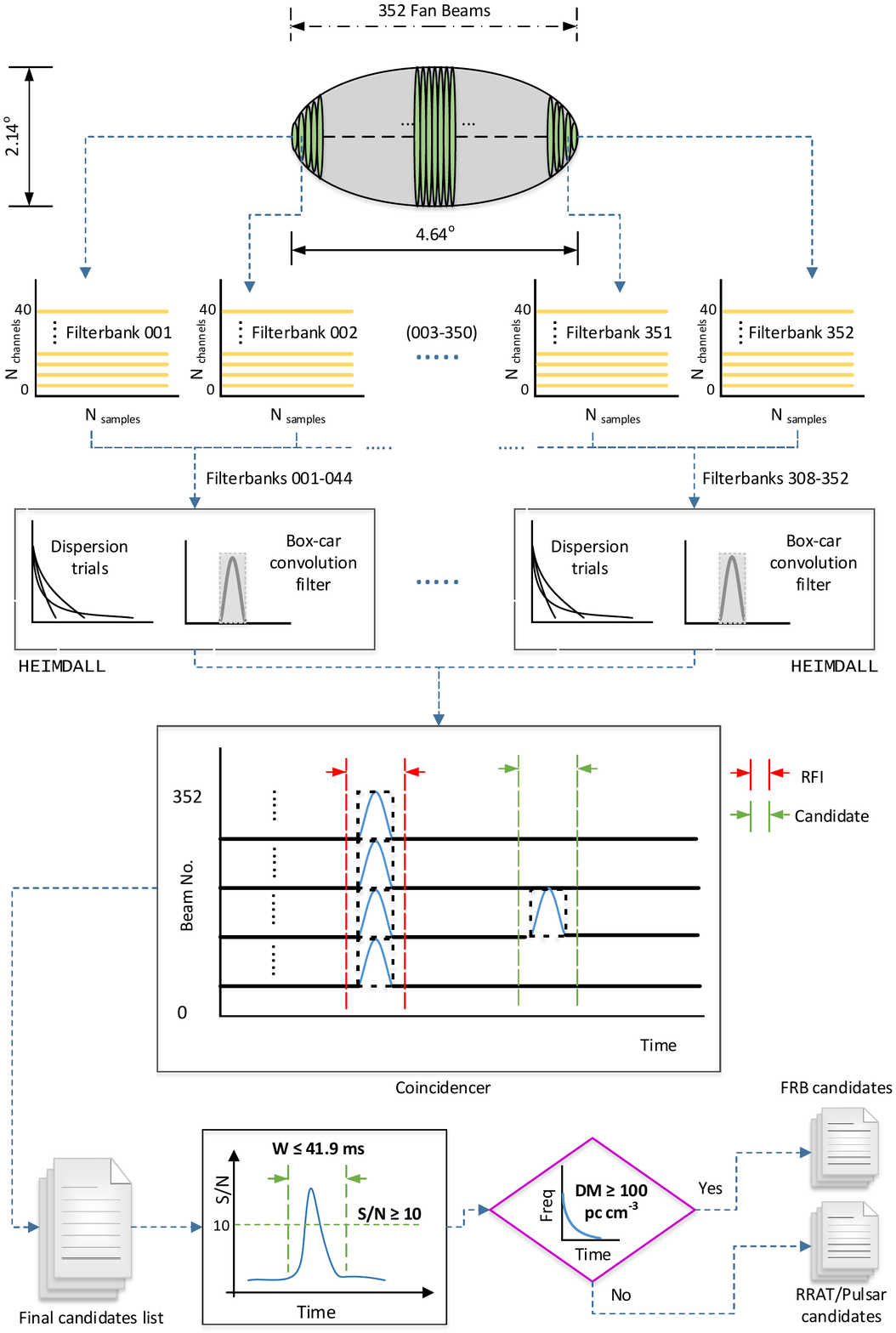}
\caption{Fast transient search pipeline at UTMOST for FRB survey V2.0.}
\label{fig:pipeline}
\end{figure*}

For both surveys the time-frequency data for each fan beam is initially dedispersed to trial 
DMs in the range 0 to 2000 $\mathrm{pc\,cm^{-3}}$. Each dedispersed time series is then
convolved with a series of boxcar filters to maximise sensitivity to single pulses and optimised for 
processing on a GPU 
using \textsc{heimdall\footnote{http://sourceforge.net/projects/heimdall-astro/}}. 
This package was originally designed for the program at the Parkes Observatory,
and has been suitably modified to accommodate the specifications of UTMOST.
\textsc{heimdall} produces a list of candidates for each of the 352 fan beams,
which are then carefully ``coincidenced'', by rejecting events if they occur simultaneously in more than 3 fan beams (of the 352). The output of the coincidencer is a final list of candidates for human inspection. The candidate list is then further filtered to only retain events which have S/N $\geqslant 10$ (to minimise the false positive rate) and $W \leqslant$ 41.943 ms ($W$ = 2$^\mathrm{N} \times 0.65536$ ms, where N = 0,1,2...). For the purposes of labeling these as either RRAT, pulsar or FRB candidates we define all pulses with $W \leqslant$ 41.943 ms and DM $\geqslant 100$ $\mathrm{pc\,cm^{-3}}$, as FRB candidates, the rest as RRAT/pulsar candidates.

\begin{figure*}
\includegraphics[width=5.85 in]{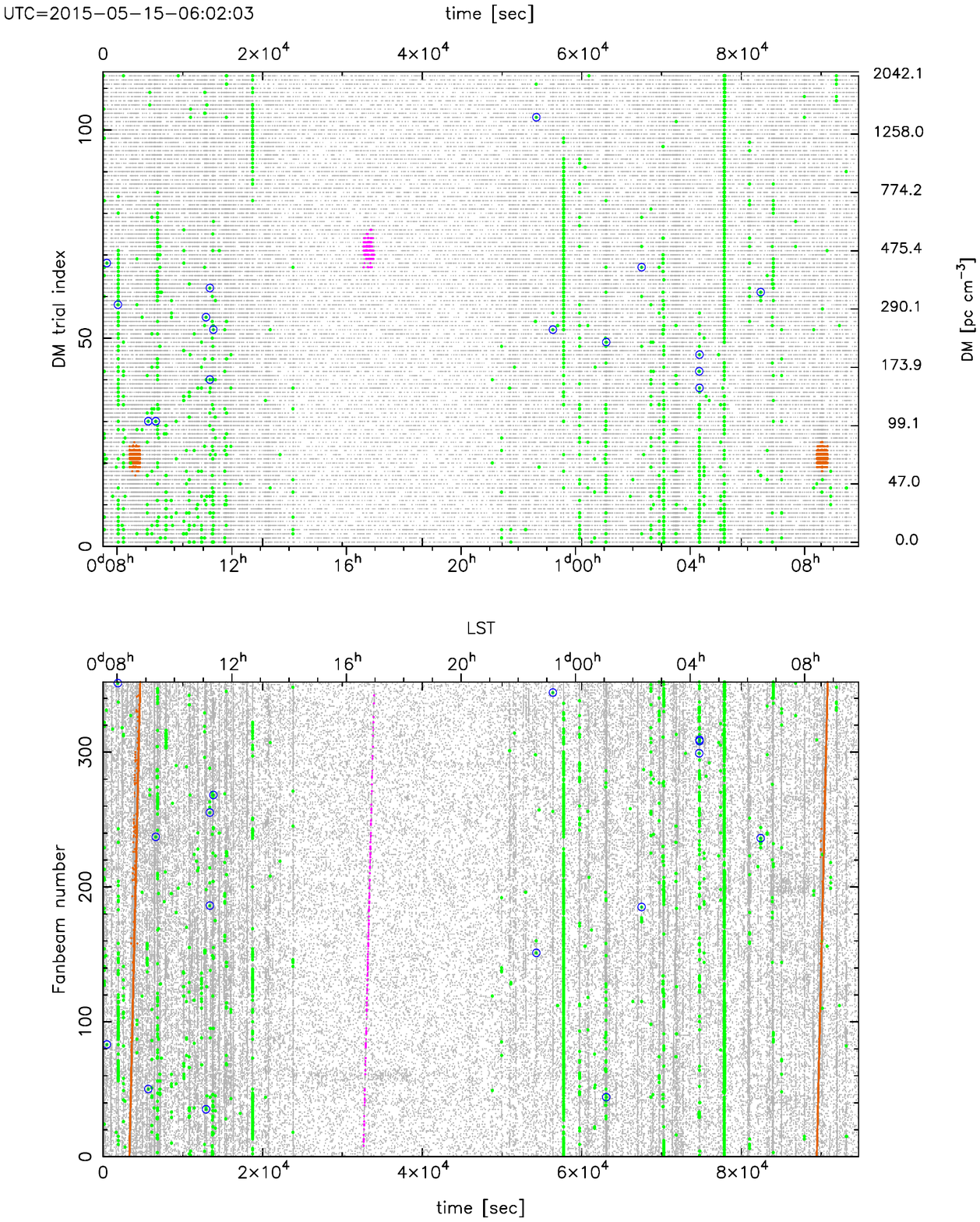}
\caption{FRB transit search from survey V1.0 at UTMOST spanning 1.1 days, starting at UTC 2015-05-15-06:02:03. The telescope was parked on the meridian at $\delta = -46^{\circ}$
Detections of the pulsars Vela and PSR J1644$-$4559 are shown by the orange and magenta points
respectively. The lines are slanted at the sidereal rate as the objects
pass through the fan beams on the sky. Events with a S/N $<$ 10 are marked in grey,
and FRB candidates by blue circles. Events with a S/N $>$ 10 but which
can be removed because they occur in 4 or more fan beams at the
same time are marked in green. These are dominated by mobile phone calls in our observing band. Indicative DM values, for the DM trial indices are shown on the right side of the upper panel in $\mathrm{pc\,cm^{-3}}$. All the FRB candidates turned out to be false positives from mobile phone calls, due usually to 20 ms narrow band emission. Events have been ignored if the total number seen in 10 second blocks exceeded 500 -- this removes about half the events but only affects about $3\%$ of the survey time-on-sky.}
\label{fig:coinc}
\end{figure*}

A typical FRB search is made on the transiting sky, with the
telescope being parked on the meridian at a declination of
$\delta = -46^{\circ}$. This declination has the advantage of the bright
Southern Hemisphere pulsars PSR J0835$-$4510 (Vela) and PSR J1644$-$4559 transiting through
the beam once per sidereal period, as well as a bright unresolved phase 
calibrator, the radio galaxy J1935$-$4620, so that the phases and delays on the array can be
checked every 24 hours. In practice, the array remains well
phased over a few days, and the calibrator was merely used used as a confirmation
of phase stability. Individual pulses from both Vela and PSR J1644$-$4559 were routinely
detected with each transit. The single pulses from PSR J1644$-$4559, with its rather high DM (478.8
$\mathrm{pc\,cm^{-3}}$) and widely spaced pulses ($P0 =$ 455 ms), and
average pulse fluence (29 Jy ms) have rather similar properties to
FRB pulses, making it an excellent daily validation of the system performance.
Figure \ref{fig:coinc} shows the passage of Vela (orange) and
PSR J1644$-$4559 (magenta) through the 352 fan beams across the sky. 
Typically a few hundred candidates would be produced per 24 hours, 
and further analyses of these candidates is performed to look for
FRBs.
The vast majority of the events in the search were RFI due
to mobile phone handsets, which operate on 5 MHz bands in our
frequency range.  The false positive rate is of order $10^{2}$ events per 12 hours across all beams in transit mode. Tests have demonstrated that coincidencing is a very
efficient means of rejecting RFI, primarily because UTMOST is an interferometer. Additionally techniques of spectral kurtosis and total power thresholding have been implemented
to mitigate the RFI. The spectral kurtosis approach measures the similarity between the input signal and Gaussian noise.
RFI typically is non-Gaussian and so this is useful in discriminating between the two (Bailes et al. in prep). The total power technique monitors and measures the median and median standard deviation for the preceding 8 seconds of data to determine when RFI causes the power levels to exceed pre-defined limits. The two techniques complement each other and result in relatively robust excision of transient RFI in an otherwise noisy environment.

Encouragingly, we redetected the
bright pulsar PSR J1430$-$6623 in our FRB search V1.0, when the telescope was
erroneously left surveying for 24 hours at its declination ($\delta = -66^\circ$) in a true blind test of system performance. Pulse recovery tests were also performed by injecting fake FRBs by adding in total power at random positions into real filterbank data. The filterbank data chosen was RFI affected and contained bright pulses from the Vela pulsar. The fake events had injected S/Ns in the range 10 to 40, DMs in the range 300 to 2000 $\mathrm{pc\,cm^{-3}}$ and widths in the range 1 to 20 ms. Blind single pulse search techniques were used to process this fake data, identical to the method used for `real' data processing. The injected FRBs were recovered with a success rate of order 95\%. All the injected FRBs with relatively high S/Ns (S/Ns $\gtrsim 15$) were re-dectected successfully and only a few with S/N = 10 were missed due to being amidst RFI. 

\begin{figure*}
\centering
\includegraphics[width=5.5 in]{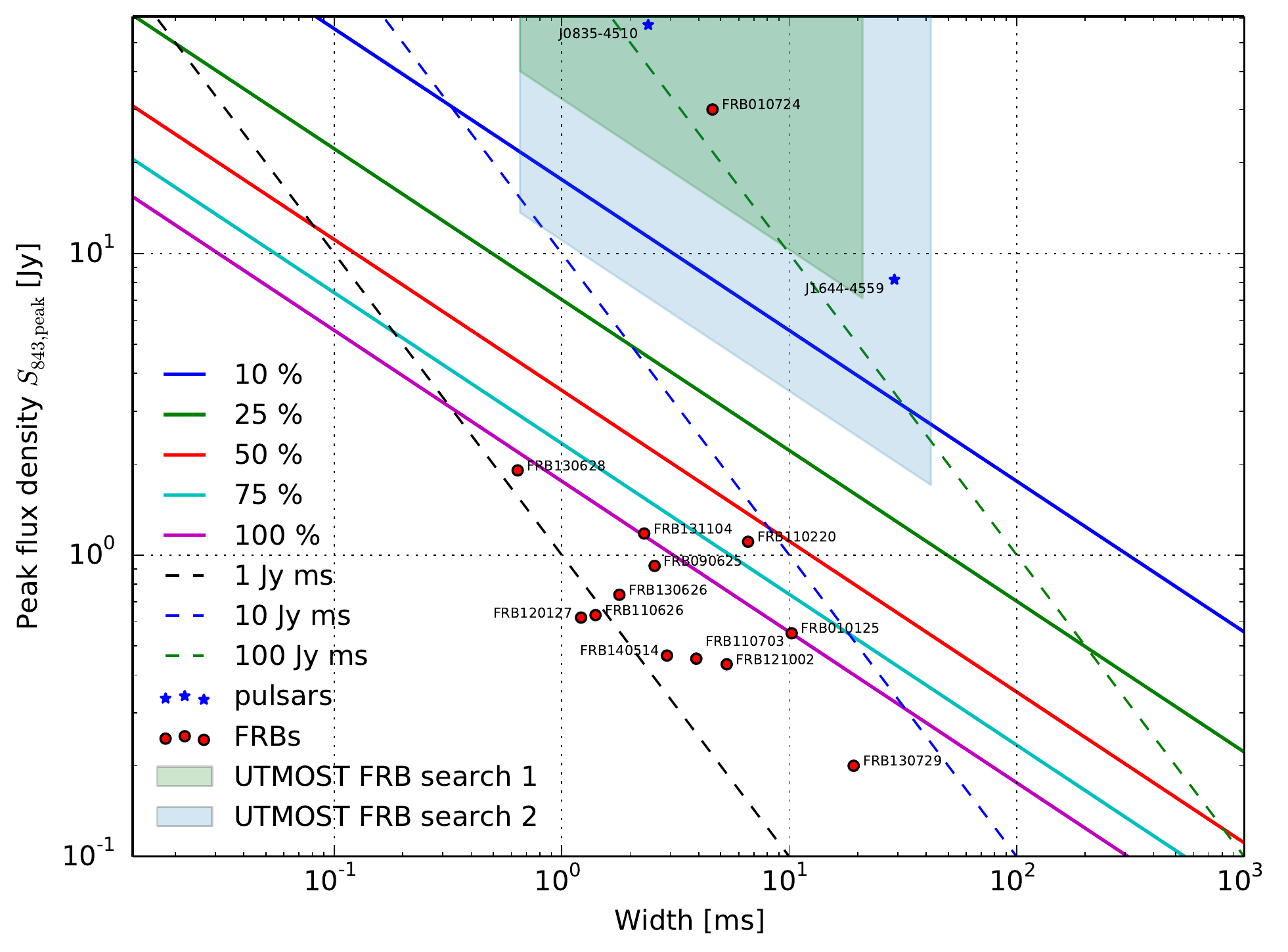}
\caption{Single pulse sensitivities at UTMOST for FRB searches V1.0 (April 2015) and V2.0 (September 2015). Solid lines of constant S/N and dashed lines of constant fluence for different fractional sensitivities of the telescope are shown. Filled circles mark the FRBs and stars mark pulsars that have been detected through their single pulses. The green and blue shaded regions enclose the UTMOST surveys at the S/N limit for the sensitivities during the V1.0 and V2.0 surveys respectively.}
\label{fig:Fig3}
\end{figure*}

\subsection{FRB Survey V1.0}

By April 2015, our upgrade of UTMOST had reached a sensitivity where we could perform an
FRB survey as part of commissioning science. FRB searches commenced, covering the full UTMOST primary beam area of 7.80 $\mathrm{deg^{2}}$. 
The system was operated at only a fraction ($\sim$ 7\%) of its final
sensitivity, as not all modules had been recommissioned, only half the
bandwidth was available (16 MHz of a final system
bandwidth of 31.25 MHz ; with 20 coarse channels), and the individual modules were still in the
process of being brought to full sensitivity. The data processing described in Section \ref{sec:survey} involving dedispersion and box-car convolution by the \textsc{heimdall} single pulse search software was performed offline by processing one filterbank at a time on a single GPU.
We typically see of order 100 pulses from PSR J1644$-$4559, with single pulse S/Ns of $\sim$ 20, when it transits the search area. 
From S/N measurements of the correlation amplitude of the quasar 3C 273 used to phase the array, we estimate $T_\mathrm{sys}$ = 400 $\pm$ 100 K. This $T_\mathrm{sys}$ yields a single pulse sensitivity of 23 $\pm$ 6 Jy ms for a millisecond duration event in 
the V1.0 search with UTMOST for a S/N of 10, using a gain of 1.4 K $\mathrm{Jy^{-1}}$ for 140 modules and a bandwidth of 16 MHz. 
We have attempted to verify the $T_\mathrm{sys}$ from transits of Vela, J1644$-$4559, J1731$-$4744 and J1752$-$2806. The uncertain flux densities yield a rather poor $T_\mathrm{sys}$ constraint with an uncertainty of a factor of 2. We set the $T_\mathrm{sys}$ = 400 $\pm$ 100 K for this survey. 
At the time of FRB survey V1.0 the telescope was at about 7\% of its design sensitivity. The search is thus sensitive to the brightest FRB reported to date referred to as a `Lorimer' type burst \citep{Lorimer}, but not yet to the brightest of FRBs reported in \cite{Thornton}. 
Assuming a Euclidean Universe, so that the cumulative number density of detectable events scales as $\mathcal{F}^{-3/2}$, we obtain an event rate estimate of about 300 days with an error margin of 50\%. With such a low sensitivity, our expectation of discovering an FRB was very low, but the survey allowed us to do many validation measurements on the FRB search pipeline.
Our total search time on sky was 467 hours. No FRBs were detected down to a fluence of 23 Jy ms and a lower S/N limit of 10. Figure \ref{fig:Fig3} displays this region (shaded in green) surveyed by UTMOST in survey V1.0. Assuming a 2$\sigma$ upper limit of 4 events \citep{Gehrels} on this null detection, 467 hours
on sky and a search area of 7.80 $\mathrm{deg^{2}}$, this yields a 2$\sigma$ upper
limit on the FRB rate at UTMOST, of not more than 1000 events $\mathrm{sky^{-1}}$ $\mathrm{day^{-1}}$ at 843 MHz with a fluence greater than 23 Jy ms.

\subsection{FRB Survey V2.0}

In September 2015, we roughly doubled our search sensitivity by doubling the number of commissioned modules. In addition the installation of the second part of the GPU correlator in May 2015 enabled us to process 31.25 MHz in 40 coarse channels. $T_\mathrm{sys}$ remained at 400 $\pm$ 100 K. The single pulses from the pulsar J1644$-$4559 were once again used to validate the system performance.
On average we found the S/N of an individual pulse to be $\sim$ 40, a factor of 2 increase from the search V1.0, i.e. the searches were at about 14\% of the design sensitivity. As previously mentioned, even though we are able to process the full 31.25 MHz bandwidth in this search, only 16 MHz of usable bandwidth resulted due to a sharp roll-off at the edges of the bandpass. The 10$\sigma$, 1 ms single pulse sensitivity during this search was 11 $\pm$ 3 Jy ms for a gain of 3.0 K$\mathrm{Jy^{-1}}$, bandwidth of 16 MHz and $T_\mathrm{sys}$ of 400 $\pm$ 100 K.

\begin{figure*}
\centering
\includegraphics[width=5.3 in]{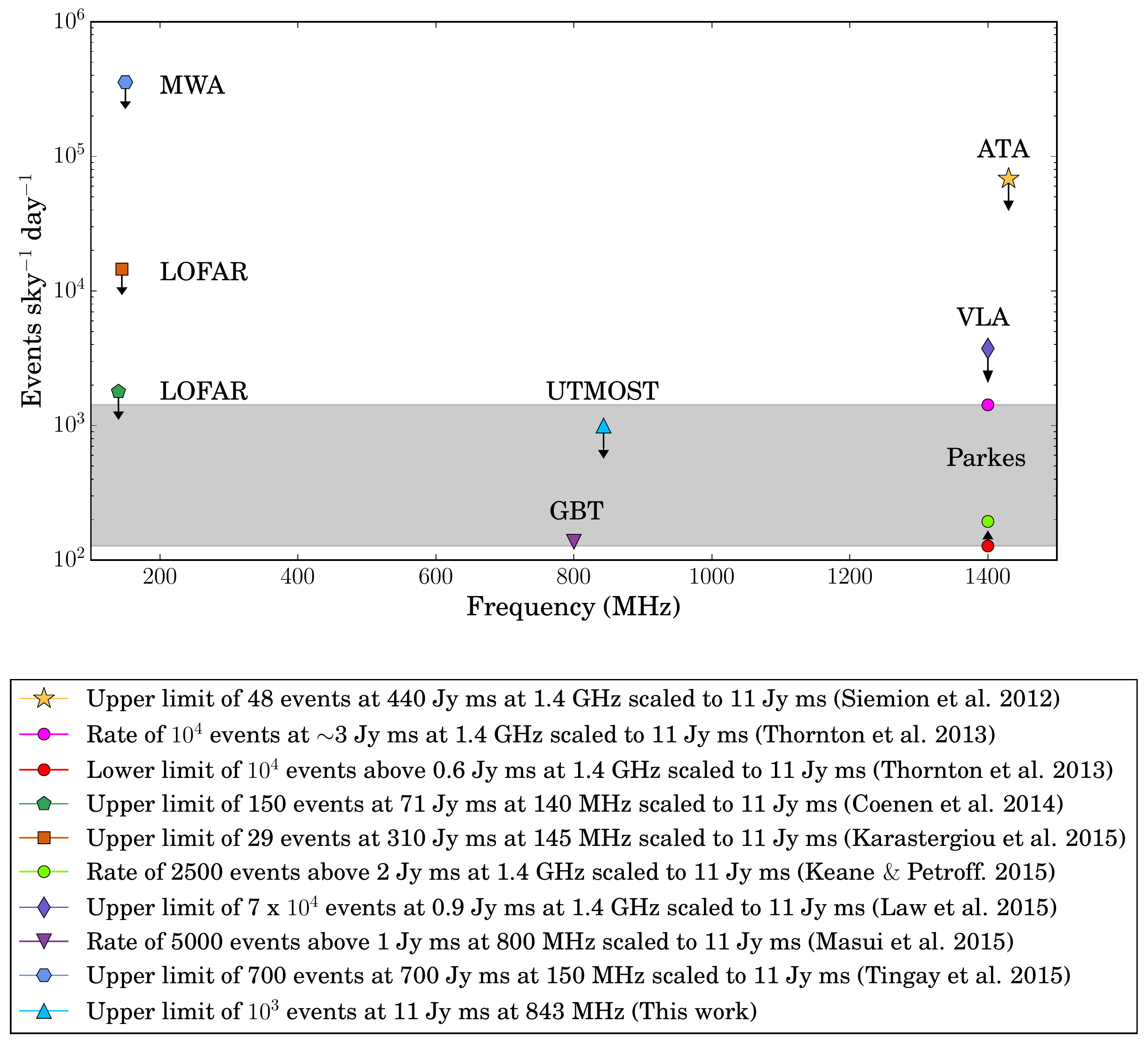}
\caption{Comparison of the non-detection of UTMOST with the published rate limits at different frequencies scaled to 11 Jy ms. Using the scaled rate of $\sim$ 200 events $\mathrm{sky^{-1}}$ $\mathrm{day^{-1}}$ at 1.4 GHz down to a fluence of 11 Jy ms (green circle) and our 2$\sigma$ upper limit of 1000 events $\mathrm{sky^{-1}}$ $\mathrm{day^{-1}}$ at 843 MHz (blue triangle), we set a lower limit on the mean spectral index of FRBs of $\alpha > -3.2$ over this frequency range.}
\label{fig:rates}
\end{figure*}

The FRB survey V2.0 was performed simultaneously with a pulsar timing programme. Additional coincidencing was performed by rejecting RFI induced events if they occurred in groups of 500 or more in 10 second intervals across all beams. The processes of dedispersion and box-car convolution by \textsc{heimdall}, was performed in real-time by processing all 352 filterbank files in blocks of 44 on 8 GPUs. 
This increased sensitivity has enabled us to detect single pulses from several more pulsars.
The RRAT J1819$-$1458 ($S_\mathrm{peak} = 3.6$ Jy at 1.4 GHz) was also detected during commensal observations with Parkes at 1.4 GHz, and UTMOST at 843 MHz. 
Figure \ref{fig:Fig3} displays the area surveyed (blue and green) by UTMOST during the V2.0 FRB search with increased sensitivity. From Figure \ref{fig:Fig3} we see that we are still only sensitive to `Lorimer' type bursts. We spent 225 hours on sky and detected no FRBs down to the fluence limit of 11 Jy ms. From this null detection we obtain a 2$\sigma$ upper
limit of not more than 1000 events $\mathrm{sky^{-1}}$ $\mathrm{day^{-1}}$ at 843 MHz with a fluence greater than 11 Jy ms.We estimate the rate of FRBs given our current sensitivity at about 120 days with an error margin of 50\%. 
Our non-detection is consistent with published FRB rate limits at Parkes, the Very Large Array (VLA) and the Allen Telescope Array (ATA) at 1.4 GHz, the Green Bank Telescope (GBT) at 800 MHz, the Low Frequency Array for Radio Astronomy (LOFAR) at 145 MHz and the Murchison Widefield Array (MWA) at 150 MHz (Figure \ref{fig:rates}) assuming the Euclidean scaling and 1 ms duration pulses.
For comparison, the current estimated rate at 1.4 GHz at Parkes is $\sim$ 200 events $\mathrm{sky^{-1}}$ $\mathrm{day^{-1}}$ down to a fluence of 11 Jy ms (green circle in Figure \ref{fig:rates}) based on the rate of 2500 events $\mathrm{sky^{-1}}$ $\mathrm{day^{-1}}$ above 2 Jy ms by \cite{KeanePetroff}. Using this scaled rate estimate at 1.4 GHz and our 2$\sigma$ upper limit of 1000 events $\mathrm{sky^{-1}}$ $\mathrm{day^{-1}}$ at 843 MHz, we estimate the FRB spectra to be no steeper than a spectral index of $-3.2$ assuming FRBs were occurring during the duration of our observations.
We have also included the \cite{Thornton} rate of $1.0^{+0.6}_{-0.5}\times10^4$ events $\mathrm{sky^{-1}}$ $\mathrm{day^{-1}}$ at 3 Jy ms (magenta circle in Figure \ref{fig:rates}) and a lower limit of $\sim$ 130 events $\mathrm{sky^{-1}}$ $\mathrm{day^{-1}}$ at 0.6 Jy ms (red circle in Figure \ref{fig:rates}) which is the event with the lowest fluence in \cite{Thornton}. The rate of $5 \times 10^3$ events $\mathrm{sky^{-1}}$ $\mathrm{day^{-1}}$ above 1 Jy ms at 800 MHz from \citep{Masui} scaled to our 11 Jy ms fluence limit is $\sim$140 events $\mathrm{sky^{-1}}$ $\mathrm{day^{-1}}$.
From LOFAR at 145 MHz we estimate not more than $\sim$ 1800 events $\mathrm{sky^{-1}}$ $\mathrm{day^{-1}}$ down to 11 Jy ms (green pentagon in Figure \ref{fig:rates}) based on the \citep{Coenen} upper limit of 150 events $\mathrm{sky^{-1}}$ $\mathrm{day^{-1}}$ brighter than 71 Jy ms. We obtain another upper limit with LOFAR at 145 MHz of $\sim 1.5\times 10^{4}$ events $\mathrm{sky^{-1}}$ $\mathrm{day^{-1}}$ down to a fluence of 11 Jy ms (orange square in Figure \ref{fig:rates}) based on the \cite{KarastergiouFRB} upper limit of 29 events $\mathrm{sky^{-1}}$ $\mathrm{day^{-1}}$ at 310 Jy ms assuming standard cosmological scaling for a fluence limited survey (viz. Section \ref{sec:scaling}). The upper limit at 1.4 GHz at the VLA is $\sim 3.7 \times 10^{3}$ events $\mathrm{sky^{-1}}$ $\mathrm{day^{-1}}$ based on $7 \times 10^4$ events $\mathrm{sky^{-1}}$ $\mathrm{day^{-1}}$ on 0.9 Jy ms events (300 mJy events of 3 ms width - their Figure 9) \citep{Law}
and at the ATA is $\sim 7 \times 10^{4}$ events $\mathrm{sky^{-1}}$ $\mathrm{day^{-1}}$ based on an event rate of 48 events $\mathrm{sky^{-1}}$ $\mathrm{day^{-1}}$ above 440 Jy ms events \citep{Siemion} scaled to the 11 Jy ms sensitivity at UTMOST. For the MWA we estimate a rate of not more than $\sim 3.5 \times 10^{5}$ events $\mathrm{sky^{-1}}$ $\mathrm{day^{-1}}$ down to a fluence of 11 Jy ms based on the upper limit of 700 events $\mathrm{sky^{-1}}$ $\mathrm{day^{-1}}$ at 150 MHz brighter than 700 Jy ms \citep{Tingay}.

\section{Estimates of FRB rates at UTMOST}
\label{sec:MR}

\subsection{Extant estimates}

\cite{Hassall} have estimated FRB rates that might be seen at a wide
range of radio telescopes operating over a wide range of
frequencies. They assumed the bursts to be standard candles, to have a constant spectral index and a
constant co-moving space density. They estimated a detection rate of
$\sim$ 3 per day at MOST, but this is an overestimate for the
present system being installed. The MOST telescope
specifications they adopt from \cite{Green} are for a more ambitious
upgrade path than the current UTMOST design, which has a bandwidth
a factor of 6 smaller and a field of view smaller
by 60$\%$ (Bailes et al. in prep). We now estimate the rate of FRBs for the current upgrade at
UTMOST using two methods, in both cases scaling from the event rate at
Parkes.

\subsection{Empirical scaling from events at Parkes}
\label{sec:empirical}

We compute what fraction of the Parkes FRBs would be detectable at
UTMOST given its sensitivity for a pulse of S/N = 10 and $W =$ 1 ms is ${S_\mathrm{min}} =$ 1.6 Jy compared to ${S_\mathrm{min}} =$ 0.4 Jy at Parkes. \cite{Thornton} discovered 4 FRBs in 24$\%$ of the high latitude (Hilat) sub-survey of the high time resolution Universe survey and estimated a rate of $1.0^{+0.6}_{-0.5}\times10^4$ events $\mathrm{sky^{-1}}$ $\mathrm{day^{-1}}$. \cite{Champion} have discovered 5 more FRBs in the remaining 75$\%$ of the survey. Thus 9 of the known FRBs were discovered in the Hilat survey alone.
To estimate their detectability at UTMOST, we reduce the S/Ns of these 9 FRBs by a factor of 4 due to UTMOSTs lower sensitivity and account for reduction in their S/N by width-broadening 
due to the difference in observing frequency between Parkes and UTMOST. Only one Parkes event is found to be 
detectable at UTMOST.

The 9 FRBs at Parkes were discovered after processing 100$\%$ of Hilat. This corresponds to a rate of $R_{\rm Parkes} = 0.08 \pm 0.03$ events $\mathrm{day^{-1}}$. Since only one of the nine Parkes FRBs is detectable at UTMOST, and the field of view is larger by a factor of 14, this yields an event rate estimate at UTMOST of $R_{\rm UTMOST} = 0.11 \pm 0.09 \, \mathrm{events} \, \mathrm{day^{-1}}$.

\subsection{Event rate based on surveyed volume}
\label{sec:scaling}

Following \citep{Hassall} we assume FRBs have flat spectral indices and are distributed through
space in a Euclidean Universe, so that the cumulative number
density of events with fluence $\mathcal{F}$, scales as $\propto
\mathcal{F}^{\alpha}$ where $\alpha = -3/2$. Since the sensitivity of
UTMOST is $\sim 0.25$ times that of Parkes, the events are expected to occur
$4^{3/2}$ = 8 times less often at UTMOST than at
Parkes. Since the area surveyed at UTMOST by the beam is 14
times that of Parkes, the overall rate is 2 times higher. Thus a rate
of 0.08$\pm$0.03 events $\mathrm{day^{-1}}$ at Parkes scales to a rate $R_{\rm UTMOST} =$
0.16$\pm$0.06 events $\mathrm{day^{-1}}$. Caleb at al. 2015 (MNRAS submitted), show using Monte Carlo simulations of a cosmological population of FRBs, that the log$N$-log$\mathcal{F}$ relation for the Parkes
events has a slope $\alpha \sim -1.0$, not as steep as the standard $\alpha =
-3/2$ relation --- adopting this shallower relation would elevate the
rate at Molonglo, but we prefer to be conservative and use $\alpha =
-3/2$, to estimate the UTMOST detection rate.

\section{Discussion and Conclusions}
\label{sec:conc}

The discovery of FRBs has opened up numerous exciting possibilities for the exploration of the extragalactic Universe. However their extragalactic/celestial origin is yet to be decisively established. With the newly upgraded UTMOST array, we will be able to affirm if these sources are truly extraterrestrial when we detect one, as the array's Fresnel zone is at $\sim$ 14,000 km. FRB searches at two different fractional sensitivities (7$\%$ and 14$\%$) were performed as part of commissioning science with the telescope parked at at $\delta = -46^{\circ}$. The chosen 
declination was to allow the diurnal passage of bright southern pulsars Vela and PSR J1644$-$4559 and a 
bright calibrator, radio galaxy J1935$-$4620, so that the system performance, phases and delays could be monitored.
No FRBs were detected down to fluence limits of 23 Jy ms and 11 Jy ms after spending 467 and 225 hours on sky respectively. 

We estimate FRB rates at UTMOST by scaling from the observed events at Parkes and by assuming a Euclidean flux distribution. Most importantly we assume a flat spectral index for both methods. The rates from the two methods at full sensitivity:  0.11$\pm$0.09
\,$\mathrm{events} \, \mathrm{day^{-1}}$ and 0.16$\pm$0.06 events
$\mathrm{day^{-1}}$, are consistent within their uncertainties. We
have used the simplest assumption sets one can apply to yield a rate at
UTMOST which shows that at full sensitivity it will detect FRBs at twice the rate as at Parkes, and yet more effective still because of our near 24$\times$7 access to the telescope. Given the duty cycle of Parkes and the estimated event rate, only a fully dedicated survey will be capable of detecting FRBs in sufficient numbers for interesting science to be performed. UTMOST is potentially the FRB discovery machine which will do this.

\section*{Acknowledgements}
The authors would like to thank the staff at the Molonglo Observatory for the exceptional support provided.
The Molonglo Observatory is owned and operated by the University of Sydney with support from 
the Australian Research Council (ARC) and the Science Foundation within the School of Physics. The 
upgrade to the telescope has been supported by the University of Sydney, Swinburne University
of Technology and  the  ARC, including via CAASTRO. Parts of this research were 
conducted by the Australian Research Council Centre for All-Sky Astrophysics (CAASTRO), 
through project number CE110001020.


\bibliographystyle{mnras}
\bibliography{bib}


\end{document}